\documentstyle[prl,aps,epsf,floats]{revtex}

\begin{document}
\draft

\twocolumn[\hsize\textwidth\columnwidth\hsize\csname @twocolumnfalse\endcsname

\title{
Bistable molecular conductors with a field-switchable dipole group
}

\author{P.\,E.\,Kornilovitch, A.\,M.\,Bratkovsky, and 
R.\,Stanley Williams}
 
\address{
Hewlett-Packard Laboratories, 1501 Page Mill Road, Palo Alto, 
California 94304
}

\date{June 24, 2002}
\maketitle

\begin{abstract}

A class of bistable ``stator-rotor" molecules is proposed, where a
stationary bridge (stator) connects the two electrodes and facilitates 
electron transport between them.  The rotor part, which has a large 
dipole moment, is attached to an atom of the stator via a single sigma 
bond.  Hydrogen bonds formed between the rotor and stator make the 
symmetric orientation of the dipole unstable.  The rotor has two 
potential minima with equal energy for rotation about the sigma bond.  
The dipole orientation, which determines the conduction state of the 
molecule, can be switched by an external electric field that changes 
the relative energy of the two potential minima.  Both orientation of
the rotor correspond to asymmetric current-voltage characteristics that 
are the reverse of each other, so they are distinguishable electrically.  
Such bistable stator-rotor molecules could potentially be used as parts 
of molecular electronic devices.

\end{abstract}

\narrowtext
\vskip2pc]

\section{Introduction}
\label{sec:one}

The success or failure of molecular electronics (moletronics) 
\cite{moletronics_one,moletronics_two} will largely depend on the 
development of efficient molecular electric switches.  Switches 
are critical to the ability to store digital information and route 
signals in moletronic logic circuits.  Single-molecule switches 
(SMS), in which the switching originates in the physical properties 
of individual molecules rather than molecular complexes or films,
are especially valuable.  SMS will allow the ultimate miniaturization 
of moletronic devices, where one memory or configuration bit is 
represented by only one molecule.

For memory applications, the SMS has to be bistable.  In principle,
the bistability may be realized as two different {\em electronic} states 
of the molecule.  Then the information can be read out by direct 
sensing of the charge on the molecule or by measuring its electrical 
conductance.  A strong dependence of conductance on the molecular  
charge state was reported by several groups \cite{Gittins,Park,Liang}.  
However, two electronic states are unlikely to have the same 
energy and the energy barrier between the states will be low.
In a pure electronic mechanism, the higher energy state can only be 
stabilized by a third electrode, a solid state gate or electrolyte, that 
affects the interior of the molecule and actually performs the switching 
between the two states.  While three-terminal switches could potentially 
be used for signal gain and routing, memory applications will be limited.  
Given the small physical dimensions of molecules, the provision of a third 
electrode to every memory bit will be difficult to achieve. 

Recently, it was suggested that a bit be stored in the form of {\em current} 
flowing through a molecule \cite{CurrentSwitch}.  Under special 
circumstances (attractive correlations between electrons on the molecule, 
and a high degeneracy of the molecular orbital), molecules could be
electronically bistable.  This is characterized by two very different 
currents passing through the molecule at the same applied voltage 
depending on the bias history.  The higher-current state is stable 
{\em only} at an external bias voltage exceeding some threshold.  
Thus this mechanism could be the basis only for volatile molecular 
memory, even if such special molecules are found.   

Alternatively, bistability may be realized as two different
{\em conformational} states of the same molecule.  In this case, the two 
states differ by the spatial positions of one or several atoms.  The 
electric conductance of the two states can be different because:
(i) the change in shape causes the current-carrying molecular orbitals
rehybridize, thus changing their energy and electron transmission; or
(ii) the redistribution of electrostatic charge significantly alters
the current-carrying orbitals through direct Coulomb interactions. 
An additional constraint is switchability in a two-terminal geometry,
which implies some degree of ionicity of mobile groups.  At least 
one class of SMS that satisfies all these requirements has been reported 
\cite{Collier_one,Collier_two,Collier_three}.  These are interlocked 
supermolecular complexes, catenanes, rotaxanes, and pseudorotaxanes,
in which the mutual orientation of two structural subunits of the
complex have at least two local energy minima.  The electrical switching 
observed in solid-state devices is speculated to be caused by the movement 
of a positively charged (+4) tetra-pyridinium ring [the full chemical name 
is ``cyclobis(paraquat-{\em p}-phenylene'')] back and forth along the
molecular backbone.  The movement happens when one of the ``stations'' on 
the backbone is oxidized or reduced under bias and pushes the ring away or
attracts it back.  Significant current hysteresis has been observed in
some cases,
although direct experimental evidence for the ring movement inside the 
devices is still lacking.  This example of ionic molecular switching is 
not the only possible mechanism.  Its main drawback is the slow switching 
speed, which is about $10^{-3}$s, due to the large mass of the moving ring. 
One way to improve the performance of SMS is to design conformationally 
bistable molecules with smaller moving parts.

\section{A novel design for a single-molecule switch}
\label{sec:two}

In this paper, we propose a new class of bistable conducting 
molecules that could be the basis for single-molecule switches, memory bits 
and other moletronic devices.  The bistability is provided by the formation 
of one or more hydrogen bonds between the moving and stationary parts of 
the molecule.  The two states are distinguishable by their current-voltage 
characteristics.  The molecules can be flipped between the two states with 
an external electric field, so that only two electrodes are required to 
operate the device.

\begin{figure}[t]
\begin{center}
\leavevmode
\hbox{
\epsfxsize=7.8cm
\epsffile{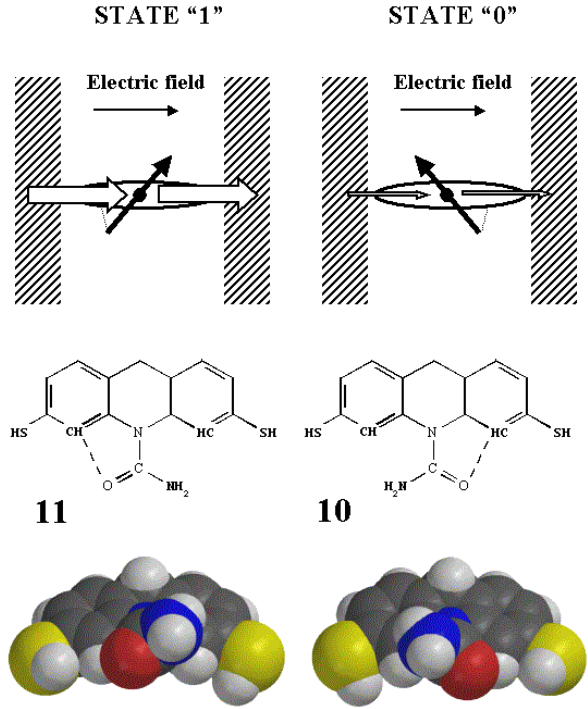}
}
\end{center}
\vspace{0.0cm}
\caption{ 
Top: The operating principle of the stator-rotor single-molecule
switch. The stator is the thin oval in the center. The dipole group,
represented by the solid arrow, is unstable with respect to one
of the two potential minima caused by the formation of hydrogen bonds
between the stator and the rotor. For electrons going from right to left,
the two states are clearly non-equivalent, which is indicated by different
widths of the open arrows. Thus the two states are distinguishable
electrically. 
Middle: A bistable stator-rotor molecule 
9-hydro-10-acridinecarboxamide-2,7-dithiol shown in state 1 ({\bf 11}) 
and in state 0 ({\bf 10}).  Hydrogen bonds formed between the oxygen of 
the amide group (-CONH$_2$) and the hydrogens of the stator are indicated 
by dashed lines. 
Bottom: Space-filling model of {\bf 11} and {\bf 10}. 
}
\label{fig1}
\end{figure}

The principal scheme of our SMS is shown in Fig.~\ref{fig1} (top).  
The molecule consists of two major subunits, the stator and the rotor. 
The stator bridges the two metal or semiconductor electrodes and facilitates 
electron transfer between them.  Thus it has to be a fairly conductive
molecular structure, which implies the stator is entirely or piecewise
conjugated.  The rotor is a side group that carries a significant dipole 
moment.  Examples of such dipole rotors are the aldehyde (-COH) and amide 
(-CONH$_{2}$) groups.  The rotor is attached to one atom of the stator via 
a single covalent sigma bond, making stator-rotor rotation relatively easy.  
The key feature of the present design is that for some stator-rotor 
molecules, the symmetric position of the rotor (that is, when it is roughly 
perpendicular to the main axis of the stator) is {\em unstable}.  This 
happens if the polar atoms of the rotor tend to form hydrogen bonds with 
the hydrogen atoms of the stator, which causes the rotor to tilt out of 
the symmetric position in either of the two possible directions.  As a 
result, the total energy of the molecule as a function of the stator-rotor 
angle has s double-well shape.  In this paper we consider the case of 
stators that are symmetric with respect to the attachment point of the rotor. 
In this case, the two equilibrium conformations must be the mirror images 
of each other.  Accordingly, the two potential minima have the same energy. 
An example of a symmetric-stator molecule is shown in Fig.~\ref{fig1} 
(middle and bottom).  The selection of one of the two equivalent potential 
minima by the rotor part can be viewed as an example of {\em discrete 
symmetry breaking.}

To complete the general description, let us mention the other two important
properties of our SMS.  Firstly, the dipole moment of the rotor interacts
with an external electric field oriented along the stator axis.  A field of 
precisely this direction is generated by a potential difference between the 
two electrodes.  Thus, a strong enough external field (external bias 
voltage) will change the energy balance between the two states, the 
population of those states becomes inequivalent and some dipolar groups 
will flip from one state to the other, performing an act of switching.  
This process is expected to be thermally-assisted.  Secondly, since both 
equilibrium states are asymmetric with respect to the electrodes, the 
current-voltage characteristics of {\em both} states will be asymmetric 
(although they will be the reverse of each other for a symmetric stator, 
as here).  Therefore, the two states can in principle be distinguished 
electrically, by applying a small test voltage (much smaller than the 
switching bias) across the electrodes and measuring the current in
one particular direction (e.g. left to right in Fig.~\ref{fig1}).
These two major properties of SMS are discussed in more detail below.

\section{Quantum chemistry calculations}
\label{sec:three}

We have performed extensive quantum chemistry calculations of several
stator-rotor molecules \cite{Spartan}. Since the results are largely similar,
we present them for 9-hydro-10-acridinecarboxamide-2,7-dithiol  ({\bf 1}) 
depicted in Fig.~\ref{fig1}.  The stator comprises three fused rings 
that can be thought of as a derivative of the fully conjugated anthracene 
molecule with one middle carbon saturated and another replaced with a 
nitrogen.  The end thiol groups -SH serve the purpose of attaching the 
molecule to gold or platinum electrodes, a standard design feature in 
moletronic studies.  The rotor is the amide group -CONH$_2$, which has 
a dipole moment of $d =$ 4.0 Debye $\approx$ 0.8 e $\cdot$ \AA.  The 
symmetric orientation of the rotor (perpendicular to the plane of the 
stator) is unstable because of the formation of {\em two} hydrogen bonds 
with the oxygen and nitrogen of the amide.

\begin{figure}[t]
\begin{center}
\leavevmode
\hbox{
\epsfxsize=8.6cm
\epsffile{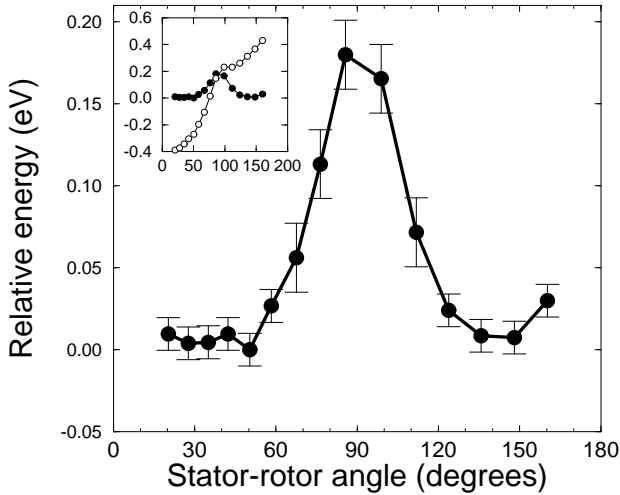}
}
\end{center}
\vspace{-0.5cm}
\caption{ 
The relative energy of the molecule shown in Fig.~\ref{fig1}, as a function 
of the stator-rotor angle.  The height of the energy barrier is 
(0.18 $\pm$ 0.02) eV. The error bars have been estimated from the energy 
variance during geometry minimization.  Inset: the same data (solid circles) 
compared with the energy of the same molecule in an external field 
of 0.5 V/\AA. 
}
\label{fig2}
\end{figure}

The relative energy of {\bf 1} as a function of the stator-rotor angle 
$\theta$, is shown in Fig.~\ref{fig2}.  It has the shape of a double well
with the height of the energy barrier $\triangle E = 0.18\pm 0.02$ eV. 
The data was obtained from complete relaxation of the molecule with the
positions of the end sulfur atoms fixed. (The latter are supposed to bind
strongly to the electrodes and therefore cannot move.)  Other useful
quantities provided by the quantum chemistry calculations are intra-molecular
vibrational modes and frequencies.  Of most interest to us is the
stator-rotor rotation mode around the N-C bond.  We identified its frequency
to be $\omega =97$ cm$^{-1}$ = 1.8 $\cdot 10^{13}$ rad/s.  This information
allows us to estimate the stability of the switch against thermal
fluctuations. The retention time can be found from Kramers formula: 
\begin{equation}
\tau =\,\frac{1}{\omega }\,e^{\triangle E/k_{B}T}.  
\label{one}
\end{equation}
Using the above values, one finds $\tau =58$ ps at room temperature 
($T=300$ K), $\tau =33$ ms at $T=77$ K, $\tau =1$ hour at $T=54$ K, and 
$\tau =30$ years at $T=40$ K.  With respect to the small retention times 
at room temperature, we should note that the energy barrier can be 
systematically increased by enhancing the hydrogen bonds and/or increasing 
their number.  For instance, replacing the -CH segments of the stator with 
-C-OH brings the hydrogens of the stator closer to the rotor and increases 
their ionicity.  As a result, the barrier increases to $0.6$ eV.  Such 
molecules, however, are more difficult to synthesize.  We believe it is 
more appropriate to focus initially on the simplest members of the 
stator-rotor family.  At the same time, we point out that ways to increase 
the energy barrier and retention times do exist in the framework of the 
present design.

Another important parameter that can be estimated from the quantum chemistry
data is the switching voltage. Taking into account the interaction energy of
the dipole moment $d$ with an external electric field $F$, the full 
$\theta $-dependent energy of the molecule becomes 
\begin{equation}
E(\theta ) = -Fd\cos {\theta } + \triangle E(\theta ),  
\label{two}
\end{equation}
where the second term represents the function plotted in Fig.~\ref{fig2}. 
At {\em zero} temperature, switching occurs when the field changes the 
energy balance of the two states such that the barrier reduces to zero and 
the system rolls down to the only energy minimum available. This is 
illustrated in the inset of Fig.~\ref{fig2}.  Equating the 
$\theta $-derivative of Eq.(\ref{two}) to zero and evaluating 
$\partial \triangle E/\partial \theta $ as finite differences, one finds 
that the barrier vanishes at field $F=0.5$ V/\AA , and the higher-energy
state of the rotor becomes mechanically unstable.  The nominal length 
of the molecule (sulfur-to-sulfur) is 10.3 \AA . Adjusting for two 
sulfur-electrode bond lengths, the distance between the two electrodes is 
about 14 \AA . This results in a switching voltage of $\approx$ 7 volts. 
At non-zero temperatures, switching occurs at a smaller voltage,
namely when the barrier is reduced such that the dipole is quickly flipped
by thermal fluctuations. One should mention that these estimates do not take
into account interaction of the molecule with other molecules or with the
electrodes.  We believe, however, that they represent the correct 
magnitude of the switching voltage.

\section{Transport calculations}
\label{sec:four}

The bistability alone is not sufficient to produce a molecular switch. 
The two states have to be distinguishable by their current-voltage (I-V)
characteristics.  We have performed self-consistent quantum-mechanical
calculations of current through molecular films of {\bf 1} sandwiched
between two semiinfinite gold electrodes.  Starting with the equilibrium
molecular structure from the quantum-chemical calculations, the end hydrogen
atoms were removed and each replaced with a triangle of gold atoms. Together
with the sulfur, the gold atoms formed a triangular pyramid with sulfur at
the pinnacle. The main axis of the molecule was near perpendicular to the
bases of the pyramids and the distance between the bases and the sulfur
atoms was 1.9 \AA \cite{Sellers}.  Then the 
(gold triangle)-molecule-(gold triangle) complexes were organized in a 
two-dimensional monolayer commensurate with the Au(111) surface and placed 
between two semi-infinite gold electrodes.  There was one molecule-gold 
complex per four surface atoms. 

The current through the monolayer of {\bf 1} was calculated from the 
B\"uttiker-Landauer formula \cite{Imry} 
\begin{equation}
I(V) = \frac{2e}{h}\int_{E_{F}}^{E_{F}+eV}dE\,T(E),  
\label{three}
\end{equation}
where $E_{F}$ is the Fermi energy of one of the electrodes, and $T(E)$ is
the transmission probability for an electron with energy $E$ to get through
the molecules from one electrodes to the other.  A tight-binding
parameterization of the molecules \cite{Harrison} and the electrodes 
\cite{Papa} as well as the details of calculation of $T(E)$ with transport 
Green's function formalism have been described elsewhere 
\cite{Sanvito,KB01,BK02,diode}.

\begin{figure}[t]
\begin{center}
\leavevmode
\hbox{
\epsfxsize=8.6cm
\epsffile{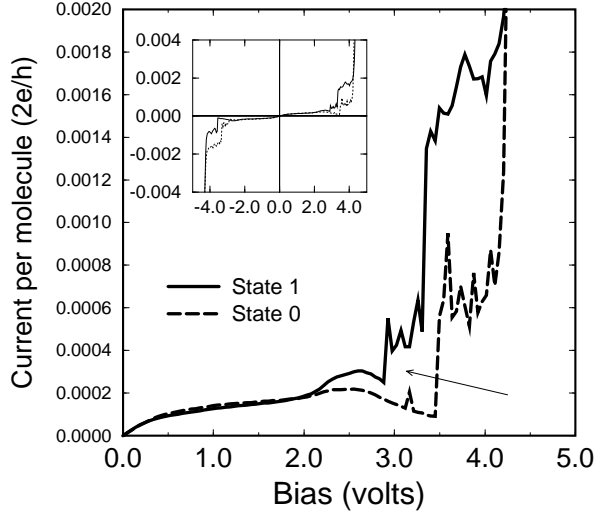}
}
\end{center}
\vspace{-0.5cm}
\caption{ 
I-V characteristics of the two states {\bf 11} and {\bf 10} of the 
stator-rotor SMS {\bf 1}.  The major ``window of distinguishability'' 
lies in the interval of 3 to 4 volts.  An arrow indicates the region where 
the currents in the two states differ because of the asymmetric 
localization of the highest occupied molecular orbital, see text.
The inset shows the same I-V characteristics in a symmetric voltage 
interval.  Note that the two curves are not exact reverses of each other.  
}
\label{fig3}
\end{figure}

The calculated I-V characteristics of the stator-rotor SMS are shown in 
Fig.~\ref{fig3}. The two curves correspond to the two stable states {\bf 11}
and {\bf 10} of the molecule {\bf 1}.  The curves are {\em not} exact 
reverses of each other because {\bf 10} was not prepared as a geometrical 
mirror image of state {\bf 11}.  Instead, the geometries of the two states 
were optimized independently.  Then the molecules were oriented 
{\em approximately} perpendicular to the surfaces of the electrodes. 
As a result, the two geometries slightly differed, which produced different 
noise in the I-V characteristics that can be seen in Fig.~\ref{fig3}. 
We emphasize in this regard that in any real situation the molecules 
would not be all locked in the same ideal conformational state but 
rather be distributed over a variety of conformations.  Thus it is 
important to investigate the sensitivity of the I-V characteristic to 
changes in the junction geometry.  We have shown earlier that the current 
through chemisorbed molecules strongly depends on their orientation with 
respect to the electrodes \cite{KB01,BK02}.  We shall discuss below that, 
given the roughness of the electrode and disorder in the film, there will 
be a spread of the I-V characteristics of individual molecules, and the 
corresponding deterioration in the performance of the molecular device.

The most important feature of the I-V characteristic of Fig.\ref{fig3} is
the hysteresis loop at voltages between 3.2 and 4.2 Volts.  Since the
currents in the two curves differ by about a factor of two, this data
demonstrates that the two states could be clearly distinguishable as long as
the current ratio remains stable and is not washed out by fluctuations of
various kinds.

\begin{figure}[t]
\begin{center}
\leavevmode
\hbox{
\epsfxsize=8.6cm
\epsffile{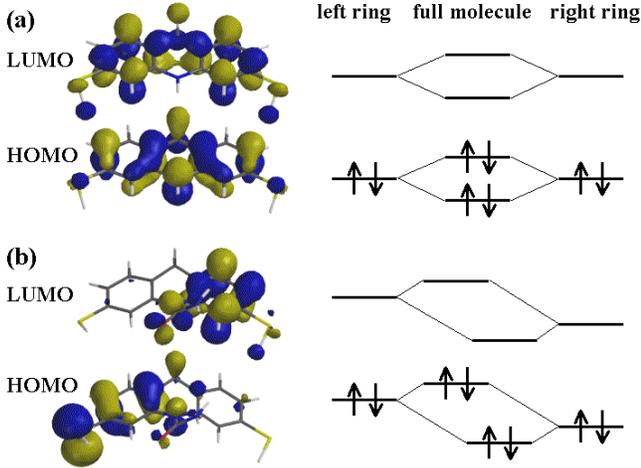}
}
\end{center}
\vspace{0.0cm}
\caption{ 
On the left: the highest occupied molecular orbitals (HOMO) and lowest 
unoccupied molecular orbitals (LUMO) of (a) relaxed stator, and (b) full 
stator-rotor molecular switch in state {\bf 11}.
On the right: the electronic structure of the molecules perceived as 
a hybridization of the electronic structures of two benzene rings. 
In {\bf 11} (b), the electric field of the dipole moment of the rotor 
shifts the levels of the left ring upward.  As a result, the HOMO (LUMO) 
of the molecule shifts toward the left (right) benzene ring. 
}
\label{fig4}
\end{figure}

We now discuss the physical origins of the observed
hysteresis. In both states, the asymmetry of the I-V characteristic must
originate from the asymmetry of molecular orbitals (MO), which in turn is
caused by the electric field of a dipole. To illustrate this point,
we compare in Fig.~\ref{fig4} the highest occupied molecular orbital (HOMO)
and the lowest unoccupied molecular orbital (LUMO) of {\bf 11}
[Fig.~\ref{fig4}(b)] with the HOMO and LUMO of the stator
alone (no dipole group) in its fully relaxed geometry [Fig.~\ref{fig4}(a)].
It is instructive to view the electronic structure of a full molecule as a
hybridization of those of the two benzene rings, see the right half of 
Fig.~\ref{fig4}. In case (a), the molecule is symmetric, the two rings are
identical, and the resulting HOMO and LUMO are equally distributed over the
two halves of the molecule. Such an electronic structure will produce a
symmetric current-voltage characteristic if the two electrodes and
molecule-electrode contacts are made the same (which we always assume in
this paper). In case (b), however, the electric field of the tilted rotor
shifts the electronic structures of the benzene rings with respect to each
other. In {\bf 11}, the electron rich (i.e. negatively charged) oxygen atom
of the rotor is closer to the left benzene ring, while the positively
charged -NH$_{2}$ group is closer to the right benzene ring. As a result,
the electronic levels of the left ring end up being higher than the right
ones by as much as 0.4 eV. After hybridization, the LUMO of the entire
molecule is primarily located on the right ring while the HOMO is on the
left ring. Clearly, such an electrical structure should produce an 
{\em asymmetric} I-V characteristics. The {\em degree} of asymmetry depends 
on the energy, conductivity, and spatial structure of the current-carrying
molecular orbitals that are closest to the Fermi level of the electrodes.

\begin{figure}[t]
\begin{center}
\leavevmode
\hbox{
\epsfxsize=8.6cm
\epsffile{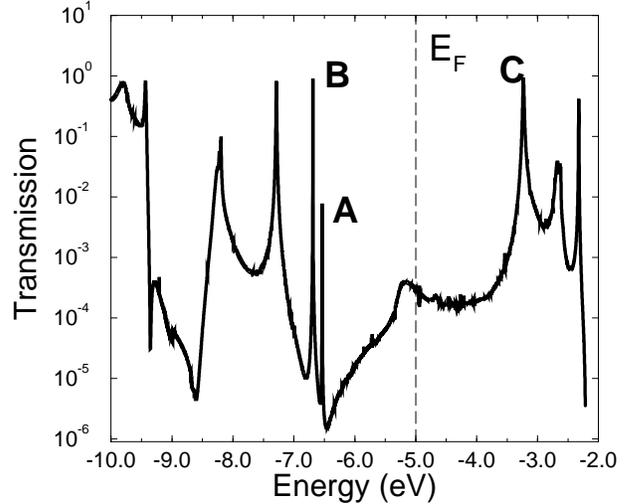}
}
\end{center}
\vspace{-0.5cm}
\caption{ 
Transmission through {\bf 11}. The dashed line indicates the
position of the Fermi level of the electrodes (-5.0 eV). 
}
\label{fig5}
\end{figure}

\begin{figure}[t]
\begin{center}
\leavevmode
\hbox{
\epsfxsize=8.6cm
\epsffile{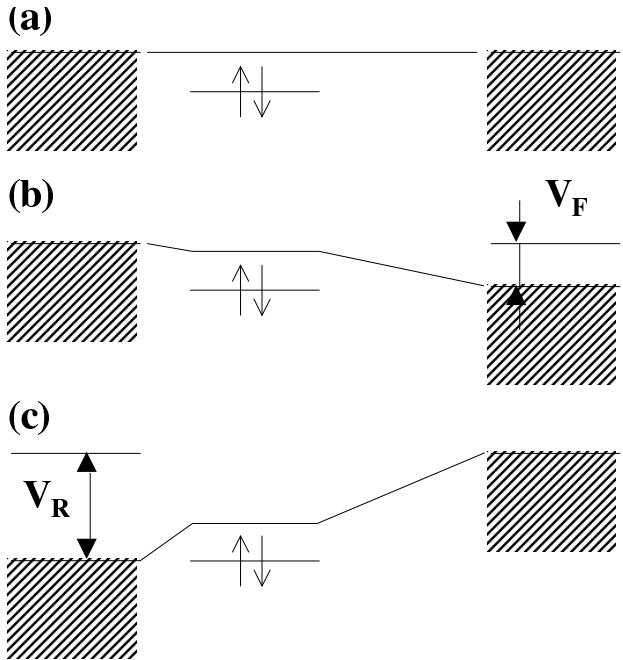}
}
\end{center}
\vspace{0.0cm}
\caption{ 
The mechanism of formation of an asymmetric I-V characteristic.
(a) A conducting molecular orbital localized closer to the left electrode
than to the right.  Most of the external bias drops on the right
molecule-electrode contact. 
(b) In the forward direction, the conditions for resonant tunneling are 
met at a relatively small bias $V_F$. 
(c) In the reverse direction, the conditions for resonant tunneling are 
met at a larger bias $V_R$. 
}
\label{fig6}
\end{figure}

The mechanism responsible for the asymmetry of the I-V characteristic can 
be understood with the help of Figs.\ref{fig5} and \ref{fig6}. 
Figure~\ref{fig5} shows the energy-dependent transmission coefficient 
$T(E)$ for {\bf 11}.  Of most interest for us are the peaks marked $A$, 
$B$, and $C$.  Peak $A$ corresponds to the HOMO.  It is the closest 
molecular orbital  
to the Fermi level (1.5 eV below) but is less conductive than $B$ or $C$. 
$B$ and $C$ are almost equidistant from $E_{F}$ by $\approx 1.75$ eV. 
It is critically important that the closest conducting level (HOMO in our 
case) is {\em spatially} asymmetric. This is the major source of the 
I-V asymmetry, as illustrated in Fig.\ref{fig6}.  When an orbital is 
shifted from the center of the molecule toward one of the electrodes, 
the total voltage drop will be distributed unevenly between the two 
contacts.  As a result, the condition for resonant tunneling will be 
met at different external biases for the two opposite polarities, 
compare (b) and (c) in Fig.~\ref{fig6}.  This simple argument applies 
equally well to the HOMO, LUMO, or any other conducting MO.  If more than 
one MO are involved, the partial asymmetries may cancel.  Thus the main 
contribution comes from the closest conducting MO.  For our SMS, the
hysteresis region caused by the HOMO is indicated by an arrow in 
Fig.~\ref{fig3}.  It turns out that this region is not the dominant one in 
the calculated I-V characteristic.  This is because of the relatively low
conductance of the HOMO, compare $A$ with $B$ and $C$ in Fig.~\ref{fig5}.

Two design rules of how to enhance the hysteresis regions of the I-V
characteristic follow from the above analysis. The {\em height} of the
hysteresis loop (that is the current difference in states ``1'' and ``0'' at 
the same voltage bias) depends on the conductivity of the MO closest to 
the Fermi level.  Indeed, the current difference appears because in {\bf 11} 
the carriers tunnel through the molecule resonantly, via a molecular level,
while in {\bf 10} the carriers must go under the barrier in non-resonant 
mode.  A strongly transmitting MO then produces a larger current difference. 
The {\em width} of the hysteresis loop depends primarily on the spatial
asymmetry of the conducting MO, as explained above. In turn, the latter
depends on the conjugation level of the stator, the length of the insulating
bridge of the stator, and the size and orientation of the dipole moment of
the rotor. Basically, the same factors determine the bistability of the
molecule, which may be affected if the molecule is changed in some way to
increase the localization of the orbital.  One general way to enhance
the asymmetry of the MO is to use an asymmetric stator.  Within the present 
design, when one wants to preserve the bistable potential profile for the 
dipole group and yet break the symmetry of the backbone, one may e.g. 
replace one or two carbon atoms in one of the side rings by nitrogen atoms.
These issues will be addressed elsewhere.

\begin{figure}[t]
\begin{center}
\leavevmode
\hbox{
\epsfxsize=8.6cm
\epsffile{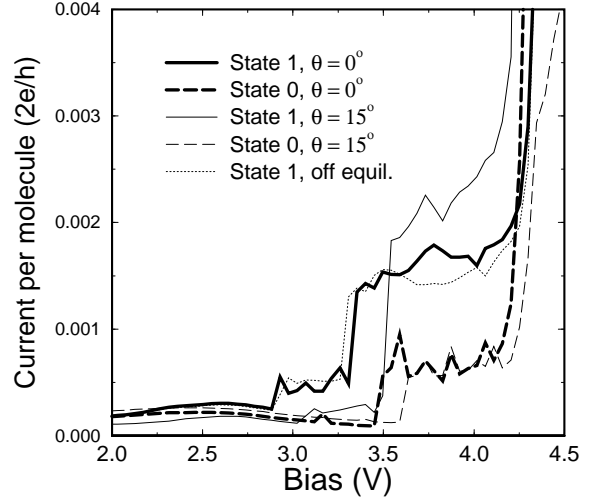}
}
\end{center}
\vspace{-0.5cm}
\caption{ 
Sensitivity of the I-V characteristic to temperature and
orientational disorder.  The thick lines are the data from 
Fig.~\ref{fig3} (the molecule is roughly perpendicular to the surface. 
The thin solid and dashed lines are the I-V characteristic of the same 
molecule but tilted by 15$^{\circ}$ from the normal (orientational 
disorder). The dotted line corresponds to state {\bf 11}, normal 
to the surface, but with the dipole swayed from its equilibrium 
conformation by 15$^{\circ}$ (temperature disorder). Orientational 
disorder seems to be more effective in closing the hysteresis region 
than the temperature one. 
}
\label{fig8}
\end{figure}

Lastly, we discuss the variation of the I-V characteristic with changing
molecular conformation and with changing geometry of the molecule-electrode
contact.  In addition to the disorder in the film, an important source of 
fluctuating geometry of the molecule is temperature.  At finite temperature, 
the rotor fluctuates about its equilibrium position which produces a 
fluctuating field on the stator and moves the conducting orbitals up and 
down in energy.  We have estimated that at room temperature the rotation
amplitude of the rotor of {\bf 1} is about 15$^{\circ }$.  Figure~\ref{fig8}
compares {\bf 11} with a rotor in its ground state and when it is
swayed away by 15$^{\circ }$. One can see that the difference in I-V
characteristics is negligible. One must add that the inelastic tunneling
through the molecules is also strongly temperature dependent and may be
observed.  This mechanism is beyond the scope of the present paper.

The prime source of the fluctuations of the contact geometry is the
nonuniformity of the electrode surfaces and molecular films. We investigated
this effect by tilting the main axis of the molecule by 15$^{\circ }$ from
the normal to the surface.  Such a tilt affects the overlap of molecular
orbitals with the wave functions of the electrode \cite{KB01}.  As a result, 
the conductance as well as the {\em position} of the resonance shift which 
might cause the closure of the hysteresis loop.  The calculated I-V
characteristics of {\bf 1} are shown in Fig.~\ref{fig8}. One can see that 
the changes relative to the normal position are quite significant. 
We conclude that the nonuniformity of the metal-molecule contact 
geometry is a serious issue that might impede the observation of the 
hysteresis and switching, unless the molecule with a wide hysteresis 
loop is chosen.

\section{Experimental tests}
\label{sec:five}

The proposed design of the stator-rotor single-molecule switch is based on 
{\em three} major effects that should work together within the same
molecular-electrode device. These are (i) the bistability of the molecules,
(ii) the distinguishability of the two states by their I-V characteristics,
and (iii) the ability to switch the molecules between the states with an
external field. It is very possible that within a particular junction, only
one or two but not all three effects will work at once. Identification of
any one of them would still be important because it would show paths to SMS
with better characteristics. We now briefly discuss the experimental modes
that can be used to approach the problem.

(i) {\em Bistability}. This is the most definite theoretical prediction for
the stator-rotor molecules. Of course, not every such molecule is bistable.
For instance, the rotor may be so large that the steric hindrance prevents it
from rotating and leaves the symmetric orientation as the only stable one.
However, for those molecules that {\em are} predicted to be bistable, the 
results are uniform across a number of modeling methods ranging from molecular
mechanics to density functional calculations. Thus while the quantitative 
numbers
(the barrier height, etc) might carry some uncertainty, the very fact of
bistability is robust.  The question is whether the bistability is detectable
by simpler than moletronic means, such as optical methods.
Consider our SMS {\bf 1} from Fig.\ref{fig1}.  In the gas phase, the two 
states would be the {\em exact} mirror images of each other and their 
infrared (IR) spectra would be indistinguishable. Now, suppose the 
molecules self-assemble on a gold surface such that they bind with one 
thiol group only. The presence of the surface will break the symmetry 
and make the two states inequivalent. As a consequence, some of the IR 
peaks will split. Moreover, the two states will be split in energy by 
$\sim $ 10-100 K (the scale of dipole-dipole interaction on separation 
1-2 nm). Thus the partial intensities of the IR doublets are expected to 
have a strong temperature dependence.

(ii) {\em Distinguishability}. It is possible that a molecule is bistable,
the energy barrier is low, yet the electric field is too weak to switch it.
This may be because of the long length of the molecule, low breakdown
threshold, effective screening, and so on. In this case, it will still be
possible to detect the current differences between states ``0'' and ``1'', 
where the switching of molecules between the states will be thermally 
activated.  A signature of such behavior would be telegraph noise 
observed at a {\em fixed} bias voltage between $V_{F}$ and $V_{R}$. 
The frequency of the telegraph noise would be strongly (exponentially) 
dependent on temperature and the field. (In interpreting the data, one 
will have to take into account other possible sources of telegraph noise, 
like the diffusive motion of metal atoms near the molecules at the 
molecule-electrode contact.) 

The described approach would be difficult to realize on molecular films
because there will be significant fractions of molecules in states ``0'' 
and ``1'' at any given time and the net average would not fluctuate much. 
Instead, measurements should be performed on single molecules.  It would 
also eliminate the disorder-induced washing out of the hysteresis regions
discussed in the previous section.  A number of experimental techniques
allowing single-molecule electrical measurements have been developed in
recent years \cite{Gittins,Park,Liang,Bumm,Reed,Lindsay,Schoen,Reichert}.

(iii) {\em Switchability}. It may happen that the molecule is bistable and
switchable by an external electric field, but does not possess a large
hysteresis loop because of the weak influence of the dipole on the
current-carrying orbitals of the stator. Then, in principle, the switching
can be detected by other means such as optical. One possibility is to shine
light through one of the electrodes that is IR transparent or made thin
enough to let some light through. In this case, the electrodes should
probably be made of different materials to make the two states of the switch
inequivalent.

Obviously, the best demonstration would be the direct observation of a I-V
hysteresis, preferably in the single-molecule measurement.

\section{Summary}
\label{sec:six}

In this paper, we have proposed a class of bistable stator-rotor molecules
that could be used for single-molecule switches and possibly for other
moletronic applications. The basic idea is to incorporate into the molecule
a rotor with a large dipole moment that performs {\em three} roles
simultaneously. Firstly, the rotor is capable of forming hydrogen bonds with
the stator, which makes the entire molecule bistable. These two stable
conformational states of the rotor can, in principle, represent the digital
``zero'' and ``one''. Secondly, the rotor makes the two states
distinguishable electrically. In either state, the rotor is positioned
asymmetrically with respect to the stator, see the bottom of Fig.\ref{fig1}.
The electric field of the dipole then lifts the mirror symmetry of the
electronic structure of the stator.  As a result, the response of the stator 
to an external bias becomes asymmetric.  Thus, the state of the switch can 
be read out by interrogating it at a small test voltage that is always the 
same in polarity and magnitude.  The two states are distinguished by the 
current passing through the junction.  Thirdly, the rotor provides the 
means to switch the molecule between the states.  The dipole moment of the 
rotor interacts with a large external electric switching field, which 
changes the relative energies of the two states and enables the molecules 
to transfer into the lowest one.  Provided that the switching voltage 
(that is, the writing voltage) is higher than the reading voltage, the 
entire read-write cycle can be done with only two external terminals.

The rotor performing three simultaneous functions results in an 
economical design of the molecular switch.  The candidate molecules are 
likely to be as small as three fused benzene rings and relatively simple 
to synthesize.  In fact, several stator-rotor molecules, similar to {\bf 1} 
but without the thiol clips, {\em are} commercially available \cite{Aldrich}. 
What is needed are the experimental methods that would allow them to be
placed between electrodes and test their conformational and electrical 
properties.

Building in and optimizing three functions within one molecule is a
challenging task. However, the proposed switching mechanism applies to a
whole class of stator-rotor molecules, rather than to one particular molecular
species. There are several ways to manipulate the mechanical and electronic
properties of the molecule by changing the chemical composition of the
stator and rotor. For instance, increasing the dipole moment of the rotor
decreases the switching voltage.  Enhancing the insulating bridge in the
middle of the stator increases the asymmetry of the I-V characteristics, and
using an asymmetric stator-rectifier widens up the hysteresis loop in I-V
curve, etc. We have also identified several experimental tests that could be
used to screen out and isolate only one or two functions (see the previous
section). One proposal is to compare infrared spectra of stator-rotor
molecule in the gas phase and self-assembled on a surface. If the molecule
is bistable, then some IR peaks should split when the molecules are on
the surface and their relative intensity should be strongly
temperature-dependent.

Finally, we mention that only symmetric stators have been discussed in this
paper. Stator-rotor molecules with {\em asymmetric} stators are richer in
physical content and potential moletronic applications. Such molecules will
be discussed in a separate publication.

\vspace{1.cm}

We thank A.S. Alexandrov and Shun-chi Chang for useful discussions.  
This work has been supported by DARPA.

\end{document}